\documentclass[11pt]{article}
\usepackage{moriond,epsfig}

\def\Journal#1#2#3#4{{#1} {\bf #2}, #3 (#4)}

\def\NIMA{{\em Nucl. Instrum. Methods} A}

\def\be{\begin{equation}}
\def\ee{\end{equation}}
\def\bea{\begin{eqnarray}}
\def\eea{\end{eqnarray}}

\begin{document}
\vspace*{4cm}
\title{NEUTRAL CHARM DECAYS AT CLEO: SEARCHES FOR {\em CP} VIOLATION AND 
MIXING}

\author{ R.M. HANS \\
	for the CLEO Collaboration}

\address{University of Illinois, Loomis Lab, 1110 W. Green St.,\\
Urbana, IL 61801 USA}

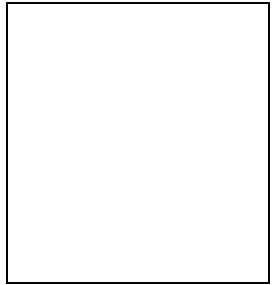
\maketitle\abstracts{Recent CLEO results on neutral charm meson decays 
presented at the XXXVIth Rencontres de Moriond are discussed.  
We find no evidence of {\em CP} asymmetry in five different 
two-body decay modes of the $D^0$ to pairs of light pseudo-scalar mesons.
We present a measurement of the mixing parameter
$y_{CP} = -0.011 \pm 0.025 \pm 0.014$
by searching for a lifetime difference between the {\em CP} 
neutral $K^+ \pi^-$ final state and the 
{\em CP} even $K^+K^-$ and $\pi^+\pi^-$ final states.  
Finally, we describe the first measurement of the rate of wrong-sign 
$D^0 \rightarrow K^+ \pi^- \pi^0$ decay:
$R_{WS} = (0.43^{+0.11}_{-0.10} \pm 0.07)\%$.
}

\section{Introduction and Motivation}
The CLEO Collaboration, operating at the Cornell Electron Storage Ring, has
been in existence for over 20 years.  We discuss recent results on neutral 
charm meson decays from a data set of $9.0$ fb$^{-1}$ of symmetric $e^+ e^-$ 
collisions at $\sqrt{s} \approx 10.6$ GeV.
One analysis uses a larger, $13.7$ fb$^{-1}$, data set.  

The study of mixing
in the $K^0$ and $B_d^0$ sectors has provided a wealth of information
to guide the form and content of the Standard Model.  In the framework of the 
Standard Model, mixing in the charm meson sector is predicted to 
be small, making this an excellent place to
search for non-Standard Model effects.  Similarly, measurable
{\em CP} violation (CPV) phenomena in strange and 
beauty mesons are the impetus for many current and 
future  experiments.
The Standard Model predictions for CPV for charm mesons 
are of the order of $0.1\%$, with one recent conjecture of 
nearly $1\%$.  Observation of CPV in charm mesons exceeding 
the percent level would be strong evidence for non-Standard Model processes.

All of the analyses described in this paper use data collected 
with one configuration of the CLEO detector, called
CLEO II.V, except for the analysis using the larger data set mentioned above,
which also uses data collected in the CLEO II configuration.  
The CLEO detector is described in detail elsewhere.~\cite{cleo_ii}
All simulated event samples were generated using
GEANT-based simulation of the CLEO detector response.

\section{General Experimental Method}\label{genmethod}
All of the analyses presented in this paper use the same general technique,
described below, except where noted.
The $D^0$ candidates are reconstructed through the decay sequence
$D^{\star +} \rightarrow D^0 \pi^+_{\rm s}$.~\cite{charge}  
The charge of the slow pion ($\pi^+_{\rm s}$) tags the
flavor of the $D^0$ candidate at production.  The charged daughters
of the $D^0$ are required to leave hits in the silicon vertex detector
and these tracks are constrained to come from a common vertex in three 
dimensions.
The trajectory of the $D^0$ is projected back to its
intersection with the CESR luminous region to obtain the $D^0$
production point.  The $\pi^+_{\rm s}$ is refit with the requirement that
it come from the $D^0$ production point, and the confidence level of the $\chi^2$ 
of this refit is used to reject background.  

The energy release in the $D^\star \rightarrow D^0 \pi^+_{\rm s}$ decay, 
$Q \equiv M^\star - M - m_\pi$, 
obtained from the above technique is observed to have a width of 
$\sigma_Q = 190 \pm 2$ keV,~\cite{qwidth} which is a combination of
the intrinsic width and our resolution,
where $M$ and $M^\star$ are the reconstructed 
masses of the $D^0$ and $D^{\star +}$ candidates respectively, and
$m_\pi$ is the charged pion mass.  The reconstruction technique 
discussed above has also been used by CLEO to measure the $D^{*+}$
intrinsic width, $\Gamma_{D^{*+}} = 96\pm 4\pm 22$ keV 
(preliminary).~\cite{GammaD*}  
\section{{\em CP} Violation in $D^0$ Decay}
Cabibbo suppressed charm meson decays have all the necessary ingredients
for {\em CP} violation -- multiple paths to the same final state and a
weak phase difference.  
However, in order to get sizable {\em CP} violation, the 
final state interactions need to contribute non-trivial phase shifts
between the amplitudes.  Large final state interactions are a likely reason why
the prediction for the ratio of branching ratios of 
$(D^0 \rightarrow K^+ K^-) / (D^0 \rightarrow
\pi^+ \pi^-)$ yields a value roughly half of the observed
value, hence these may provide a good hunting ground 
for {\em CP} violation.

We present results of
searches for direct {\em CP} violation in neutral charm meson decay to pairs of
light pseudo-scalar mesons: $K^+ K^-$, $\pi^+ \pi^-$, $K^0_{\rm S} \pi^0$, 
$\pi^0 \pi^0$ and $K^0_{\rm S} K^0_{\rm S}$.

\subsection{Search for {\em CP} violation in $D^0 \rightarrow K^+ K^-$
and $D^0 \rightarrow \pi^+ \pi^-$ decay}

The asymmetry we want to measure,
$A = \left[ \Gamma \left( D^0 \rightarrow f \right) - 
\Gamma \left( \overline{D^0} \rightarrow f \right) \right]/$
$\left[ \Gamma \left( D^0 \rightarrow f \right) + \right.$ \\
$\left. \Gamma \left( \overline{D^0} \rightarrow f \right) \right]$
can be obtained from the asymmetry
$A^f = $$\left[ \Gamma \left( D^{\star +} \rightarrow 
\pi^+_{\rm s} f \right) - 
\Gamma \left( D^{\star -} \rightarrow \pi^-_{\rm s} f \right) \right]/$
$\left[ \Gamma \left( D^{\star +} \rightarrow \pi^+_{\rm s} f \right) +
\Gamma \left( D^{\star -} \rightarrow \pi^-_{\rm s} f \right) \right]$.
The slow pion and $D^0$ are produced by the {\em CP}-conserving strong 
decay of the $D^{\star +}$, so the slow pion serves as an unbiased flavor tag
of the $D^0$.  The decay asymmetry can be obtained from the apparent production
asymmetry shown above because the production of $D^{\star \pm}$ is 
{\em CP}-conserving.

The asymmetry result is obtained by fitting the energy release ($Q$) spectrum
of the $D^{\star +} \rightarrow D^0 \pi^+_{\rm s}$ events.  The $D^0$ mass
spectra are fit as a check.  The background-subtracted $Q$ spectrum is fit 
with a signal shape obtained from $K^+ \pi^-$ 
data and a background shape determined 
using Monte Carlo.  
The parameters of the slow pion dominate the $Q$ distribution, 
so all modes have the same shape.
We do the fits in bins of $D^0$ momentum to eliminate any biases due to
differences in the $D^0$ momentum spectra between the data and the MC.
The preliminary results are 
$A(K^+ K^-) = 0.0005 \pm 0.0218 ({\rm stat}) \pm 0.0084 
({\rm syst})$ and $A(\pi^+ \pi^-) = 0.0195 \pm 0.0322 ({\rm stat}) \pm 0.0084 
({\rm syst})$.
The measured asymmetries are consistent with zero, and no {\em CP} violation
is seen.  These results are the most precise to date.~\cite{OLDCP}

\subsection{Search for {\em CP} Violation in $D^0 \rightarrow K^0_{\rm
S} \pi^0$, $D^0 \rightarrow \pi^0 \pi^0$ and $D^0 \rightarrow K^0_{\rm S} 
K^0_{\rm S}$ decay}

This analysis~\cite{jaffe} differs from the other analyses 
presented in this paper in 
some of its reconstruction techniques and in the data set used.  
The $\pi^0 \pi^0$ and
$K^0_{\rm S} \pi^0$ final states do not provide sufficiently precise
directional information about their parent $D^0$ to use the intersection of the $D^0$ 
projection and the CESR luminous region to refit the
slow pion as described in the general experimental technique section. 
The $K^0_{\rm S} K^0_{\rm S}$ final state is treated the same for consistency.
This analysis uses the data from both the CLEO II and CLEO II.V
configurations
of the detector.

The $K^0_{\rm S}$ and $\pi^0$ candidates are constructed using only good 
quality tracks and showers.  The tracks (showers) whose combined 
invariant mass is close to the $K^0_{\rm S}$ ($\pi^0$) mass are
kinematically constrained to the $K^0_{\rm S}$ ($\pi^0$) mass, improving the 
$D^0$ mass resolution.  The tracks used to form $K^0_{\rm S}$
candidates are required to satisfy criteria designed to reduce
background from $D^0 \rightarrow \pi^+ \pi^- X$ decays and combinatorics.  
Candidate events with
reconstructed $D^0$ masses close to the known $D^0$ mass are selected to
determine the asymmetry, 
$A(f) = \left[ \Gamma (D^0 \rightarrow f) - \Gamma
(\overline{D^0} \rightarrow f)\right] /$$\left[ \Gamma 
(D^0 \rightarrow f) + \right.$ \\ 
$\left. \Gamma(\overline{D^0} \rightarrow f)\right]$.  
The total number of $D^0$ and $\overline{D^0}$ candidates for a given final
state is determined as follows.  We fit the $Q$ distribution outside of 
the signal region and interpolate the fit under the signal peak to 
determine the background in the signal region.  We subtract the background
in the signal region from the total number of events there to determine
the total number of signal events.
After background subtraction, we obtain $9099 \pm 153$ $K^0_{\rm S}
\pi^0$ candidates, $810 \pm 89$ $\pi^0 \pi^0$ candidates, and $65 \pm 14$ 
$K^0_{\rm S} K^0_{\rm S}$ candidates.  

The difference in the number of $D^0$ and $\overline{D^0}$ to a given final 
state is determined by taking the difference of the number of events in
the signal region, and the asymmetry is obtained by dividing by the
number of candidates determined above.  This method of determining the
asymmetry implicitly assumes that the background is symmetric.  

We obtain the results 
$A(K^0_{\rm S} \pi^0) = (+0.1 \pm 1.3)\%$, 
$A(\pi^0 \pi^0) = (+0.1 \pm 4.8)\%$ and 
$A(K^0_{\rm S} K^0_{\rm S}) = (-23 \pm 19)\%$ 
where the uncertainties contain the 
combined statistical and systematic uncertainties.
All measured asymmetries are consistent with zero and no indication of
significant {\em CP} violation is observed.  This measurement of 
$A(K^0_{\rm S} \pi^0)$ is a significant improvement over previous results, 
and the other two asymmetries reported are first measurements.

\section{Search for $CP$ dependent lifetime differences due to 
$D^0-\overline{D^0}$ Mixing}

In the limit of no $CP$ violation in the neutral $D$ system
we can write the time dependent rate for $D \to f$, where $f$ is a
CP eigenstate, as
$R(t) \propto e^{-t \Gamma(1 - y_{CP} \eta_{CP})}$
where $\Gamma$ is the average $D$ width, $\eta_{CP}$ is the $CP$
eigenvalue for $f$, and
$y = y_{CP} = \frac{\Delta \Gamma}{2 \Gamma}$
where $\Delta \Gamma$ is the width difference between the physical
eigenstates of the neutral $D$,
and $y$ is the standard mixing parameter.
We can then express $y_{CP}$ as
$y_{CP} = \frac{\tau_{\overline{CP}}}{\tau_{CP+}} - 1$
where $\tau_{\overline{CP}}$ is the lifetime of a $CP$ neutral state, such
as $K\pi$, and $\tau_{CP+}$ is the lifetime of a $CP$ even state,
such as $KK$ or $\pi\pi$.  Thus to measure $y_{CP}$ we simply
take the ratio of the lifetimes of $D^0 \to K \pi$ to $D^0 \to KK$ and
$\pi\pi$.  Since the final states are very similar, our backgrounds 
are small, and cross-feed among the final states is negligible, many
of the sources of uncertainty cancel in the ratio.  For a discussion of 
the case allowing {\em CP} Violation see, for example, 
reference.~\cite{petrov}

We fit the proper time distributions of the signal candidates
selected in a narrow region around the $D$ mass with an unbinned
maximum likelihood fit.
The probability for a candidate to be signal
is determined by its measured mass, and is based on the fit to the 
mass distributions.
Background is considered to have contributions with both zero 
and non-zero lifetimes.

We calculate $y_{CP}$ separately for the $KK$ and $\pi\pi$
samples.  
Our preliminary results are 
$y_{KK} = -0.019 \pm 0.029 ({\rm stat}) \pm 0.016 ({\rm syst})$
and 
$y_{\pi\pi} = 0.005 \pm 0.043 ({\rm stat}) \pm 0.018 ({\rm syst})$.
We form a weighted average of the two to get
$y_{CP} =  -0.011 \pm 0.025 ({\rm stat}) \pm 0.014 
({\rm syst}) ({\rm preliminary})$.

This result is consistent with zero, and with previous measurements of
$y_{CP}$.~\cite{ycp}

\section{First Observation of Wrong-Sign $D^0 \rightarrow K^+ \pi^- \pi^0$ Decay}

The $D^{0} \rightarrow K\pi\pi^{0}$ candidates are reconstructed using the
selection criteria described in Section~\ref{genmethod}, with additional 
requirements specific to
this analysis.  In particular, $\pi^{0}$ candidates with momenta greater than
340~M$e$V/$c$ are reconstructed from pairs of photons detected in the
CsI crystal calorimeter.  Backgrounds are reduced by requiring
specific ionization of the pion and kaon candidates to be consistent
with their respective hypotheses.

The right--sign mode was recently studied by CLEO\cite{bib:bergfeld} and found
to have a rich Dalitz structure consisting  
of $\rho(770)^{+}$, $K^{*}(892)^{-}$, $\overline{K^{*}}(892)^{0}$, 
$\rho(1700)^{+}$, $\overline{K_{0}}(1430)^{0}$, $K_{0}(1430)^{-}$,
and $K^{*}(1680)^{-}$ resonances and non-resonant contributions.
Recent theoretical predictions based on 
U-spin symmetry arguments\cite{bib:GronauRosner} suggest that the wrong sign (WS)
channel will have a different resonant substructure than the right sign (RS).  
We allow for different average WS and RS efficiencies in the
calculation of the WS rate:
$R_{WS} = (\overline{\varepsilon}_{RS}/\overline{\varepsilon}_{WS})\cdot
(N_{WS}/N_{RS})$.  

The ratio of yields is measured by performing a
maximum likelihood fit to the two-dimensional distribution in
$m(K\pi\pi)$ and $Q$.  The signal distribution in these variables is
taken from the RS data.  The
background distributions are determined using a large Monte Carlo sample,
which corresponds to approximately eight times the integrated
luminosity of the data sample.  The $Q$--$m(K\pi\pi^{0})$ fit yields a 
WS signal of 
$38 \pm 9$ events and a ratio $N_{WS}/N_{RS}=0.43^{+0.11}_{-0.10}\%$.
The statistical significance of this signal is found to be 4.9
standard deviations. 

The average efficiency ratio is determined
using a fit to the Dalitz plot variables $m^{2}(K^{+}\pi^{-})$ and
$m^{2}(K^{+}\pi^{0})$ in the WS data.  In this fit, the amplitudes and
phases are initialized to the RS values, and those
corresponding to the $K^{*}(892)^{+}$ and $K^{*}(892)^{0}$ resonances 
are floated relative to the dominant $\rho(770)^{-}$ and other minor
contributions.  Combining the square of the
fitted amplitude function with a parameterization of the efficiency
determined using a large non-resonant Monte Carlo sample, we measure
an average efficiency ratio of  $1.00\pm 0.02 ({\rm stat})$.

We measure the wrong sign rate to be
$R_{WS} = 0.43^{+0.11}_{-0.10}\ ({\rm stat}) \pm 0.07\ 
({\rm syst})\%$ (preliminary).
This measurement can be used to obtain limits on
$R_{DCSD}$ as a function of $y^\prime$
(see reference~\cite{alex}).


\section*{References}
\begin{thebibliography}{99}
\bibitem{cleo_ii} Y. Kubota {\it et al}, \Journal{\NIMA}{320}{66}{1992};
	T.S. Hill, \Journal{\NIMA}{418}{32}{1998}.

\bibitem{charge} Charge conjugation is implied throughout, except where the 
	charge conjugate states are explicitly shown, such as in an asymmetry
	definition.

\bibitem{qwidth} This result is for the $D^0 \rightarrow K^+ \pi^-$ mode.  
	Other modes have similar widths since the uncertainty on the slow pion 
	dominates the width of the $Q$ distribution.

\bibitem{GammaD*} CLEO Collaboration, T.E. Coan {\it et al.}, CLEO-CONF 01-02,
	hep-ex/0102007. 

\bibitem{OLDCP} CLEO Collaboration, J. Bartelt {\it et al.}, Phys. Rev. D 
	{\bf 52}, 4860 (1995); FOCUS Collaboration, J.M Link {\it et al.}, 
	Phys. Lett. B
	{\bf 491}, 232 (2000); Erratum-ibid. {\bf 495}, 443 (200); 
	E791 Collaboration, E.M.
	Aitala {\it et al.}, Phys. Lett. B {\bf 421}, 405
	(1998); E687 Collaboration, P.L. Frabetti {\it et al.}
	, Phys. Rev. D {\bf 50}, 2953 (1994); E691 Collaboration, 
	J.C. Anjos {\it et
	al.}, Phys. Rev. D {\bf 44}, 3371 (1991).

\bibitem{jaffe} CLEO Collaboration, G. Bonvicini {\it et al.}, Phys. Rev. D
	{\bf 63}, 071101 (2001).

\bibitem{petrov} S. Bergmann {\it et al.}, Phys. Lett. B {\bf 486}, 418 (2000).

\bibitem{ycp} E791 Collaboration, E.M. Aitala {\it et al.}, Phys. Rev. Lett.
	{\bf 83}, 32 (1999); FOCUS Collaboration, J.M. Link {\it et al.}, 
	Phys. Lett. B {\bf 485}, 62 (2000).

\bibitem{bib:bergfeld} CLEO Collaboration, S. Kopp {\it et al.}, 
	Phys. Rev. D {\bf 63} 092001 (2001).

\bibitem{bib:GronauRosner} M. Gronau, J. L. Rosner, Phys. Lett. B {\bf 500}, 
	247 (2001).

\bibitem{alex} CLEO Collaboration, G. Brandenburg {\it et al.}, hep-ex 0105002,
	submitted to Phys. Rev. Lett.

\end{thebibliography}
\end{document}